\documentclass[aps,twocolumn,pra,showpacs,amsmath,amssymb,showkeys]{revtex4-1}

\usepackage{graphicx}
\usepackage{braket}
\usepackage{hyperref}

\begin{document}

\title{Strong-coupling limit of the driven dissipative light-matter interaction}

\author{Th. K. Mavrogordatos}
\affiliation{Department of Physics, Stockholm University, SE-106 91, Stockholm, Sweden}
\email[Email address: ]{themis.mavrogordatos@fysik.su.se}

\date{\today}

\begin{abstract}
We approach the strong-coupling thermodynamic limit in the response of the open driven Jaynes-Cummings (JC) oscillator. We do so by highlighting the r\^{o}le of quantum fluctuations against the semiclassical response in three distinct regimes of operation. We begin by demonstrating the persistence of photon blockade, predicted in [H. J. Carmichael, Phys. Rev. X {\bf 5}, 031028 (2015)], as a manifestation of the inherently-quantum and nonlinear JC spectrum revealed for vanishing dissipation. We then proceed to discuss the importance of bistability, which, despite being present in photon blockade, is able to provide an alignment between the semiclassical nonlinearity and quantum dynamics only for a driving amplitude having the same order of magnitude as the light-matter coupling strength. This resolution brings us to the critical point of the well-known quantum phase transition of second order on resonance, where the quantum and semiclassical pictures are once more contrasted for a varying participation of the two coherent interactions when going through the collapse of the quasi-energy spectrum. 
\end{abstract}

\pacs{42.50.Pq, 42.50.Ct, 42.50.Lc}
\keywords{Dissipative quantum phase transitions, strong-coupling limit, photon blockade, bistability, neoclassical equations.}

\maketitle

\section{Introduction}

Critical behavior in quantum optical systems where the particle number is not conserved (see e.g. \cite{Carusotto2009, Kilda19, NoSymmetryBreaking, Baumann2010, Klaers2010}) presents substantial interest as well as conceptual challenges (for a recent discussion see \cite{CiutiSpectralTheorem} and references therein). In order to address the issue, a thermodynamic limit can be formulated via the number of system excitations for which the displayed nonlinearity responsible for criticality can no longer be treated as a negligible perturbation \cite{Carmichael2015}. Light-matter interaction in the form of the Jaynes-Cummings (JC) oscillator, the quintessential model of quantum optics \cite{JCarticle}, is subject to a limit of two kinds on the basis of the above definition, when considered in the presence of drive and dissipation. One of them is a so-called {\it weak-coupling limit}, where quantum fluctuations reduce to an inconsequential noise on top of the semiclassical response. Such a response reminds us of the laser output, in which fluctuations are some orders of magnitude lower than the mean photon occupation of the generated coherence \cite{QO1, EntanglementLaser}. On the other end of the line, this limit pertains as well to the appearance of single-atom resonant absorptive optical bistability \cite{SingleAtom}. The second one, on the other hand, is a {\it strong-coupling limit}, the one born out of a coupling strength, here between a harmonic oscillator and elemental matter in the simplest form of a two-level ``atom'', which appreciably exceeds the dissipation rate. 

The breakdown of {\it photon blockade} \textemdash{a} term coined in \cite{Concept97} \textemdash{by} means of a dissipative quantum phase transition, investigated theoretically and experimantally \cite{Carmichael2015, Fink2017, StabilityFSC}, reveals the importance of both these limits and their interplay as we explore the quantum dynamics activated by different sources of dissipation. In particular, the presence of spontaneous emission, even if very small, brings about a competing scaling behavior akin to a weak-coupling limit, gradually eroding the coherence which is manifested by multi-photon resonances and a split Lorentzian, until the whole phase portrait is significantly ``softened'' \cite{Carmichael2015}. 

Let us now look at the three interactions comprising our model. Two of them are external and one is internal. The internal interaction refers to the resonant coupling with strength $g$ of the atom with pseudospin raising and lowering operators $\sigma_{\pm}=(1/2)(\sigma_{x}\pm i \sigma_{y})$ and inversion $\sigma_z=2\sigma_{+}\sigma_{-}-1$, and the field mode with annihilation (creation) operator $a$ ($a^{\dagger}$), when these two entities exchange quanta, the familiar Rabi oscillations. One of the external interactions is coherent, namely the coupling to a (semi)classical field with a coherent state amplitude insensitive to quantum fluctuations. This field drives the JC ``molecule'' whose spectrum displays a characteristic $\sqrt{n}$ nonlinearity \cite{Fink2008, Bishop2008}. The other interaction is incoherent, via two channels of dissipation: photon loss through the cavity mirrors with rate $2\kappa$, and spontaneous emission due to coupling of the two-level system to radiation modes other than the cavity mode, with rate $\gamma$. The latter source of incoherence gives rise to a weak-coupling limit, with a system-size parameter scaling as $(\gamma/g)^2$, while the former is linked to a strong-coupling limit subject to a scaling of $(g/\kappa)^2$ \cite{Carmichael2015, SingleAtom}. In the context of these limits, what are we then to expect as the timescale determined by dissipation extends significantly to include the manifestly coherent evolution of such a fundamental quantum nonlinear oscillator? 

In this work, we will address this question by focusing on the interplay between the two aforementioned coherent interactions in the strong-coupling limit, as probed via an output channel with a vanishing dissipation rate. Spontaneous emission will be assumed absent unless explicitly stated otherwise. We will first demonstrate the persistence of photon blockade as a manifestly quantum phenomenon, although with a semiclassical flavor, where quantum fluctuations keep on generating a disagreement with the semiclassical picture due to the presence of discrete quantum energy levels against an incoherent evolution. We will then move into these regions of the drive parameter space where quantum-fluctuation switching resolves a neoclassical nonlinearity taking the form of complex-amplitude bimodality; we will find that this resolution is defined by a simple relation between the coupling strength $g$ and a particular configuration $(\Delta\omega, \varepsilon_d)$ of drive detuning and amplitude, respectively. The response here is dominated by bistability with a finite value of the strong-coupling limit scale parameter, the {\it finite-size scaling} of a first-order quantum phase transition in zero dimensions, as explained in \cite{StabilityFSC} : this is a form of instability where quantum fluctuations, owing to the uncertainty principle, cause transitions between two states with a finite lifetime exceeding appreciably the timescale set by dissipation \cite{SingleAtom}. In contrast to the quantum-fluctuation switching that we will observe in the region of quantum blockade, the standard deviation, or uncertainty, over each of the two metastable states becomes visibly smaller than the distance between their means. Finally, we will show that quantum critical behavior foreshadows symmetry breaking when we approach the organizing point of a second-order phase transition on resonance. Here, the importance of quantum fluctuations is gradually diminished as a semiclassical response takes over for $2\varepsilon_d/g \gg 1$.

In particular, after defining the equations governing the quantum and mean-field descriptions in Sec. \ref{sec:MEandNC}, we will consider the relevance of the Kerr oscillator in the development of nonlinearity, and the emergence of resonant {\it quasi} two-level multi-photon complexes in Sec. \ref{sec:photonBl} as we attain the strong-coupling thermodynamic limit. A discussion on the decoherence due to spontaneous emission in Sec. \ref{sec:spontBl} will be followed by an investigation into the regime of smaller detunings and high excitation, where we will present asymptotic expressions for the states of neoclassical bistability in Sec. \ref{sec:CABistanh}, before finally moving to resonance and a dissipative quantum phase transition of second order in Sec. \ref{sec:symmbr}. To that end, we employ semiclassical results, the solution of the master equation (ME) with exact diagonalization in a truncated Hilbert space as well as its unravelling into sample quantum trajectories (see \footnote{We use the {\it Quantum Optics Toolbox} in Matlab for exact diagonalization of the Liouvillian, and the QSD C++ library for the generation of quantum trajectories (quantum state diffusion with adaptive stepsize).} and \cite{qsdreference} for further details). 

\section{Master equation and the neoclassical scaling law}
\label{sec:MEandNC}

Within the Born-Markov approximation, the ME determining the evolution of the reduced density matrix $\rho$, namely the dynamics for which the modes of the reservoir the system is coupled to have been lumped into an effective medium with collective attributes, reads \cite{spontdressedstate, QO2}
\begin{equation}\label{MasterEq}
\dot{\rho}=[1/(i\hbar)][H, \rho]+\kappa(2a\rho a^{\dagger}-a^{\dagger}a\rho-\rho a^{\dagger}a),
\end{equation}
with the Hamiltonian in the interaction picture under the rotating wave approximation:
\begin{equation}\label{Hamiltonian}
\begin{aligned}
H=-\hbar \Delta \omega \left(\sigma_{+}\sigma_{-}+ a^{\dagger}a \right)& +i\hbar g \left(a^{\dagger}\sigma_{-}-a\sigma_{+}\right)\\
& + i\hbar\varepsilon_d \left(a^{\dagger}-a\right).
\end{aligned}
\end{equation}
Here, $\Delta\omega=\omega_d-\omega_0$ stands for the detuning, with $\omega_0$ the frequency of the resonant mode coinciding with the two-level spacing of the atom, and $\omega_d$ the frequency of the classical driving field. The first term on the right-hand side of Eq. \eqref{Hamiltonian} gives the occupation of the JC molecule, the second term refers to the coherent {\it JC interaction}, and the third term accounts for the coupling to the external drive with amplitude $\varepsilon_{d}$. The second term on the right-hand side of Eq. \eqref{MasterEq} quantifies the coupling of the resonant cavity mode to a Markovian reservoir taken at zero temperature, populating the output channel. The JC interaction produces {\it dressed states} with frequencies (in the Schr\"{o}dinger picture) $\omega_{0}=0$ and $\omega_{n,\pm}=n\omega_0\pm \sqrt{n}\,g$, $n=1, 2, \ldots$, where the two signs define two ladders of excitation, as pictured in Fig. 1 of \citep{spontdressedstate}. The coherent driving of the JC molecule on resonance ($\Delta\omega=0$) induces a further {\it dressing of the dressed states}, giving rise to the quasi-frequencies (in the interaction picture) $\Omega_{0}=0$ and $\Omega_{m, \pm}=\pm \sqrt{m}\,g[1-(2\varepsilon_d/g)^2]^{3/4}$, $m=1, 2, \ldots$ \cite{quasienergies, QO2, Carmichael2015}. The quasi-energy spectrum at resonance depends on the drive amplitude, and collapses to zero at the critical point $\varepsilon_d=g/2$. It can be extended to account for the JC-Rabi model, with an appropriate renormalization of the drive amplitude \cite{QPTJCRabi, DiracJG2018, RamanDM}.

\begin{figure*}
\begin{center}
\includegraphics[width=\textwidth]{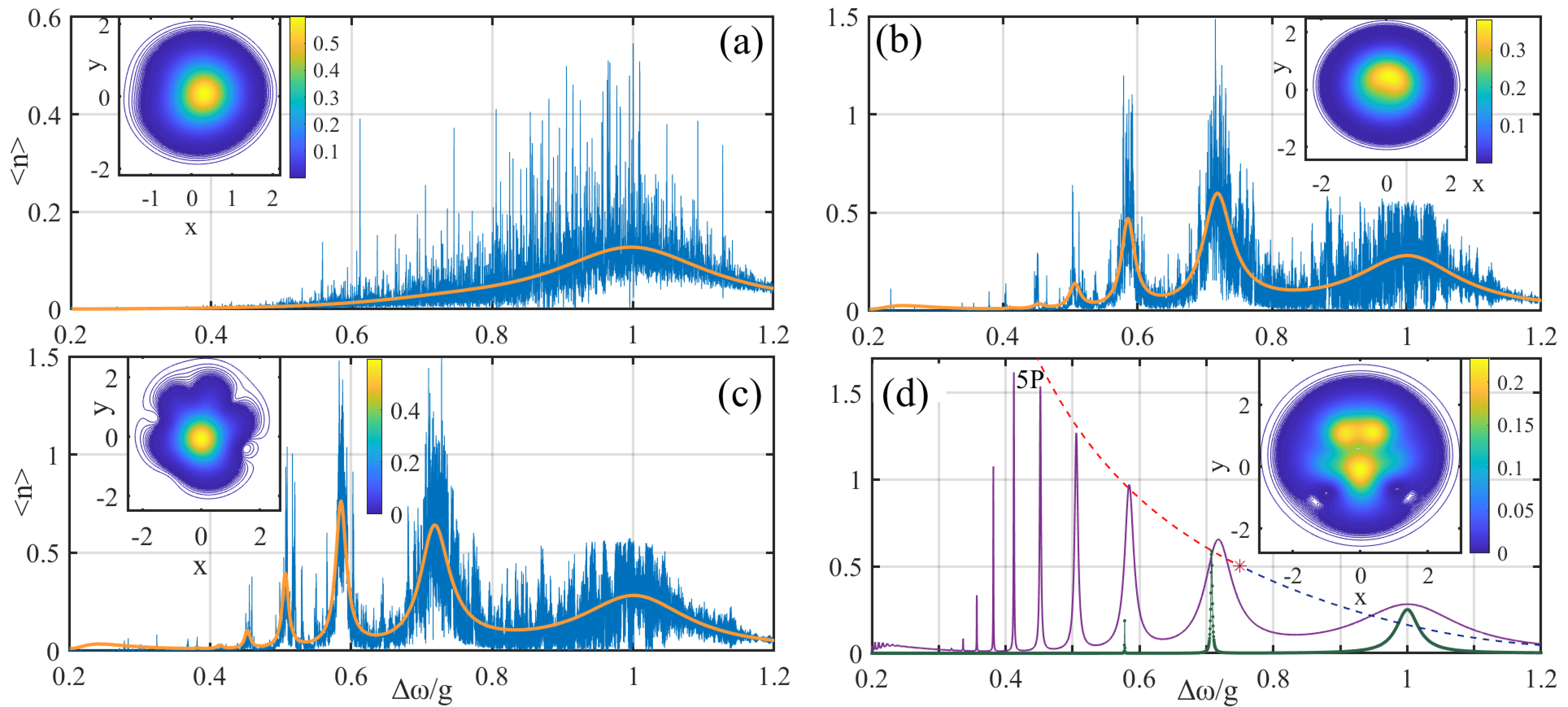}
\end{center}
\caption{{\it The persistence of photon blockade.} {\bf (a)}-{\bf (c)} Time-dependent photon number averages from single quantum trajectories for a fixed ratio $\varepsilon_{d}/g=0.09$, and time-varying detuning to coupling strength ratio $\Delta\omega(t)/g$ [with $\Delta\omega(t)/g=0.2+t/T$, $0\leq t \leq T$] for $g/\kappa=5, 100, 200$, respectively, superimposed onto the average photon number $\braket{n}_{\rm ss}\equiv\braket{a^{\dagger}a}_{\rm ss}$, as obtained from the steady-state solution of the ME \eqref{MasterEq}. {\bf (d)} $\braket{n}_{\rm ss}$ for $\varepsilon_{d}/g=0.09$ (solid purple line) and $\varepsilon_{d}/g=0.02$ (green dotted line), both with $g/\kappa=5000$. The dashed lines depict the neoclassical response for $\varepsilon_{d}/g=0.09$; past the asterisk on increasing $|\alpha_{\rm ss}|^2$, we give the upper branch of the bistability curve. The dimensionless scan time is $\kappa T=5\times 10^3$ in frame (a) and $\kappa T=5\times 10^2$ in frames (b, c). The insets in frames (a, b) and (c, d) depict contour plots of the steady-state Wigner {\it quasi}-probability distribution $W(x+iy)$ of the intracavity field for a detuning to coupling strength ratio equal to $\Delta\omega/g=1$ and $\Delta\omega/g=0.4510$, respectively. The latter value is very close to the peak of the five-photon resonance, indicated by {\it 5P} in frame (d).}
\label{fig:Blockade}
\end{figure*}
\begin{figure}
\begin{center}
\includegraphics[width=0.5\textwidth]{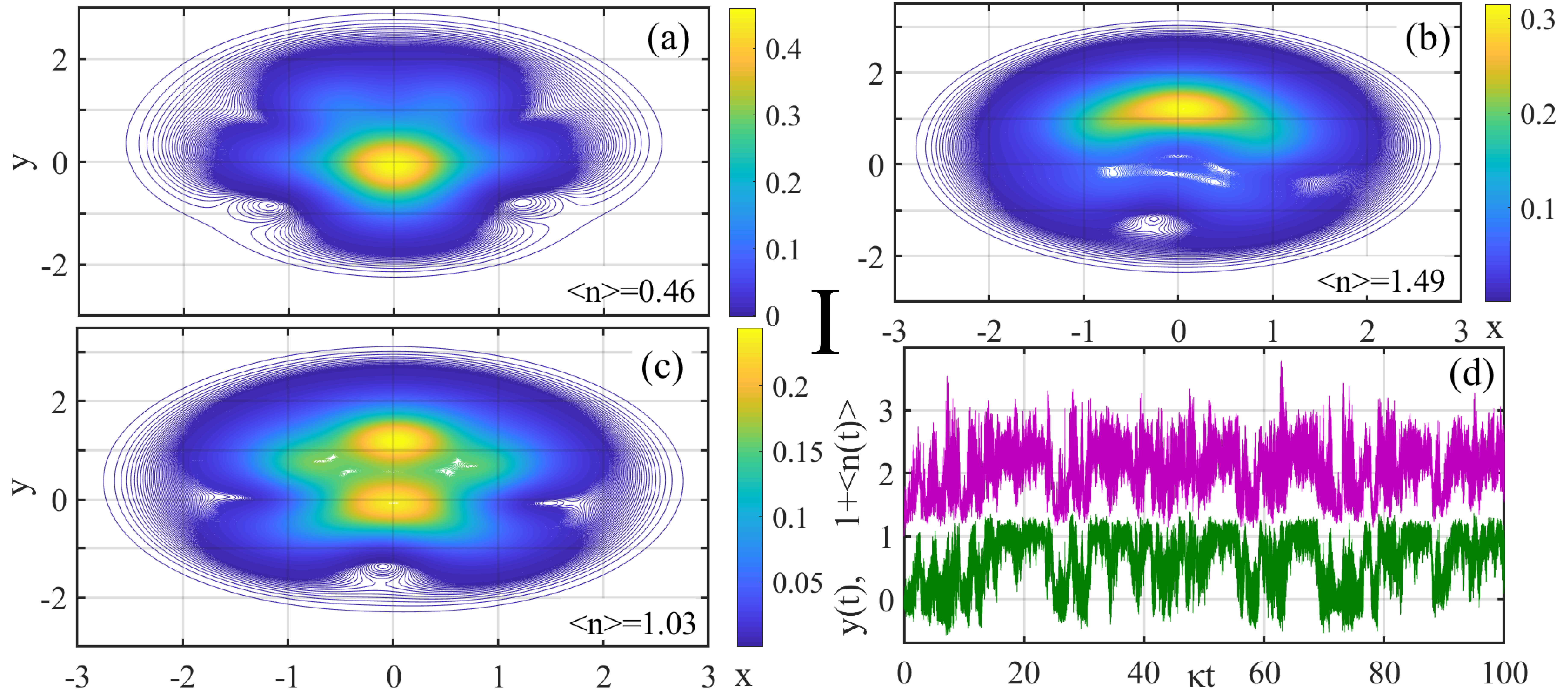}
\includegraphics[width=0.5\textwidth]{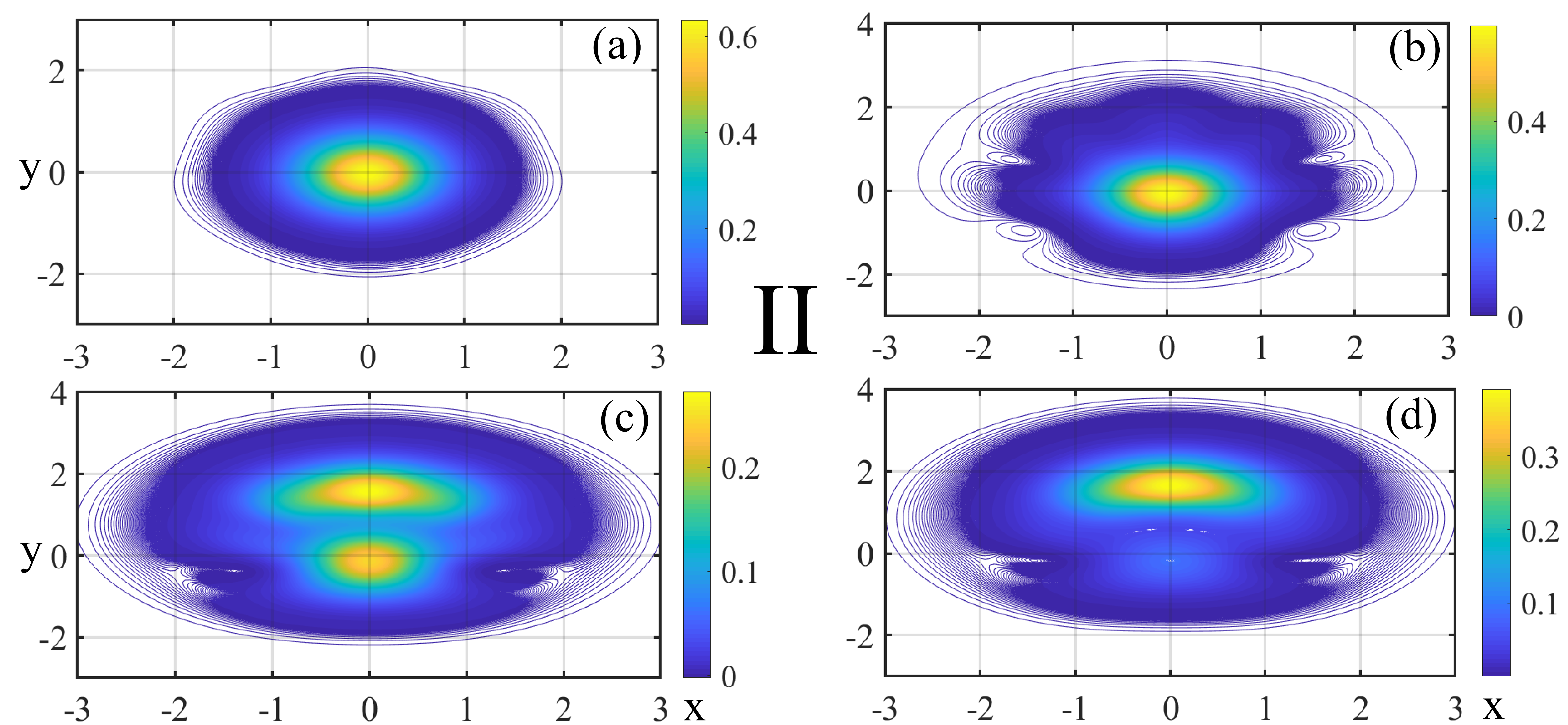}
\end{center}
\caption{{\it Scanning across an $n$-photon resonance.} \underline{Panel I}: {\bf (a)}-{\bf (c)} {\it Quasi}-probability distributions $W(x+iy)$ of the intracavity field, obtained by the steady-state solution of the ME \eqref{MasterEq}, for a fixed drive to coupling strength ratio $\varepsilon_{d}/g=0.09$ and varying detuning, tracking the five-photon resonance {\it 5P} in Fig. \ref{fig:Blockade}(d). {\bf (d)} Sample quantum trajectory depicting the imaginary part of the intracavity field and the average photon number (displaced by one for clarity), $y(t)={\rm Im}[\braket{a(t)}],\, 1+\braket{n(t)}$, respectively, for the same value of $\Delta \omega/g$ as in frame (c). In (a)-(c), $\Delta\omega/g=0.4500, 0.4520, 0.4526$, and $\braket{n}_{\rm ss}=0.46, 1.49, 1.03$, respectively. \underline{Panel II}: Wigner functions of the intracavity field, obtained by the steady-state solution of the ME \eqref{MasterEq}, for a fixed detuning to coupling strength ratio $\Delta\omega/g=1/\sqrt{6}$ and varying $\varepsilon_d/g$ as $0.048, 0.108, 0.168, 0.196$ in frames {\bf (a)}-{\bf (d)}, respectively. In both panels, $g/\kappa=5000$.}
\label{fig:5P6PWigner}
\end{figure}

Considering average values of the corresponding to the ME \eqref{MasterEq} Heisenberg equations, and factorizing the expectations of the operator products yields the {\it neoclassical} equations of motion. Solving then for the steady-state intracavity field, $\alpha_{\rm ss}$, with negative two-level inversion, we obtain an equation for its modulus, featuring the {\it strong-coupling scaling parameter} $n_{\rm sc}=g^2/(4\kappa^2)$ \cite{Carmichael2015},
\begin{equation}\label{neocl}
|\alpha_{\rm ss}|^2=(\varepsilon_{d}/\kappa)^2\,f(\Delta\omega/\kappa,g/\kappa;|\alpha_{\rm ss}|^2/n_{sc}),
\end{equation}
with 
\begin{equation*}
\begin{aligned}
f(\Delta\omega/\kappa,g/\kappa;&|\alpha_{\rm ss}|^2/n_{\rm sc})=\{1 +[\Delta\omega/\kappa -{\rm sgn}(\Delta\omega)\\
&\times (\Delta\omega^2\kappa^2/g^4 +|\alpha_{\rm ss}|^2/n_{\rm sc})^{-1/2}]^2\}^{-1},
\end{aligned}
\end{equation*}
a nonlinear function of the scaled neoclassical cavity-field amplitude. Coming from that origin, in what follows we will investigate the correspondence between the neoclassical scaling law of Eq. \eqref{neocl} and the demonstration of critical behavior in the quantum picture as we attain the thermodynamic limit in various regions of the drive parameter space $(\Delta\omega/g, \varepsilon_{d}/g)$. 

\section{From semiclassical perturbation theory to the discrete JC spectrum}
\label{sec:photonBl}

\subsection{Mapping to the Kerr oscillator}
\label{subsec:mappringKerr}

We will now track the emergence of low-amplitude nonlinearity due to the atom-field coupling, taking the form of a deviation from a Lorentzian profile. For a very small excitation, with $2g^2|\alpha_{\rm ss}|^2 \ll \Delta\omega^2$, the nonlinearity can at first be neglected, yielding a response which peaks for the two Rabi resonances at $\Delta\omega/g=\pm 1$. Further along, the first occurrence of nonlinearity is of Kerr-type, following the expansion of the square root in Eq. \eqref{neocl} to first order in $|\alpha_{\rm ss}|^2/n_{\rm K}$, with $n_{\rm K}\equiv\Delta\omega^2/(2g^2)$. A new scaling emerges for the first appearance of bistability, independent of dissipation and akin to a weak-coupling limit. In the development of the vacuum Rabi resonance with increasing $n_{\rm sc}$, for $\Delta\omega=g$ and $|\alpha_{\rm ss}|^2 \ll 1/2$, the neoclassical equation precludes any low-amplitude bistability, as the intracavity excitation reads 
\begin{equation}\label{vacRessemi}
|\alpha_{\rm ss}|^2 \approx \varepsilon_{d}^2 (\kappa^2+4g^2|\alpha_{\rm ss}|^4)^{-1},
\end{equation}
mapped to an effective Kerr oscillator with zero detuning. Taking $\kappa^2 \ll 4|\alpha_{\rm ss}|^4 g^2$ in a self-consistent fashion, we obtain $|\alpha_{\rm ss}|^2 \approx [\varepsilon_d/(2g)]^{2/3}$, valid for $\varepsilon_d/(2g) \ll 1$. When $\Delta\omega<g$ (with $\Delta\omega>0$), however, the detuning of the equivalent Kerr oscillator, $\overline{\Delta \omega}=\Delta\omega-g^2/\Delta\omega$, will eventually satisfy the condition of bistability, $(\overline{\Delta \omega})^2>3 \kappa^2$ [with $\overline{\Delta \omega}\,(g^4/\Delta\omega^3)<0$], in the limit where the dissipation rate tends to zero (see Eq. (32) of \citep{Carmichael2015} and Sec. 2 of \cite{Drummond_1980}). We note that since $n_{\rm K} \sim 1/2$ in the region of the vacuum Rabi resonance, we expect that quantum fluctuations will produce visible deviations from the mean-field nonlinearity, already for a very weak intracavity excitation. 

When the condition $2g^2|\alpha_{\rm ss}|^2 \ll \Delta\omega^2$ is not upheld, for larger excitations or for smaller detunings, the perturbative treatment of the neoclassical nonlinearity discussed above is no longer valid. Moreover, in the limit $\Delta\omega/\kappa\to 0$, a divergence is encountered for $|\alpha_{\rm ss}|\to 0$, marking the inadequacy of the semiclassical theory and the need to explicitly bring quantum levels into consideration \cite{Carmichael2015}.

In Fig. \ref{fig:Blockade}, individual quantum trajectories map out the emergence of the characteristic $\sqrt{n}$ nonlinearity through a spectrum of non-equidistant resonances for a time-varying positive detuning, with growing $n_{\rm sc}$. For a dissipation rate $\kappa$ which is of the order of the coupling strength $g$, only a broad (vacuum Rabi) resonance is present, centered at $\Delta\omega=g$ [frame (a)]. When $n_{\rm sc}$ grows, however, the single-photon transition saturates and multi-photon resonances \cite{multiphotonres} are revealed, with varying weight, some {\it pre} and others {\it post} saturation, in the driving frequency response [frames (b-d)]. The saturation of the vacuum Rabi resonance in the quantum dynamics, $\braket{a^{\dagger}a}_{\rm ss, \,max} \approx 1/4$ (see Fig. 3 of \cite{effectivetwolevel}), with a more obvious separate peak for a lower value of $\varepsilon_d/g=0.02$ and $\varepsilon_d/\kappa=100$, as shown in Fig. \ref{fig:Blockade}(d), takes us away from the validity region of the semiclassical mapping to the Kerr oscillator and the prediction $|\alpha_{\rm ss}|^2 \approx [\varepsilon_d/(2g)]^{2/3}$ of Eq. \eqref{vacRessemi} \textemdash{the} ratio $\varepsilon_d/\kappa$ must now be taken into account in the nonlinearity of saturation. 

The mean-field perturbative prediction is invalidated for a further decreasing dissipation rate and the ensuing growth of the intracavity excitation; the steady-state photon number average in the quantum picture proceeds first as a monotonically increasing and then a constant function of the ratio $\varepsilon_{d}/\kappa$ in the thermodynamic limit. The average photon number $\braket{a^{\dagger}a}_{\rm ss}$ is fixed at $1/4$ as long as $\varepsilon_{d}/g$ remains sufficiently small. Moreover, the mapping in question places the onset of bistability at 
\begin{equation}\label{DuffingDw}
\Delta\omega/g \approx \{[1+(1-\sqrt{3}\kappa/g)^2]/2\}^{1/2}
\end{equation}
in the strong-coupling limit, approaching the vacuum Rabi resonance with very low excitation; this prediction of Eq. \eqref{DuffingDw} is in contrast to what we observe in Fig. \ref{fig:Blockade}(d). Nevertheless, the heralded bistability, pertubatively of Kerr-type, will remain of relevance in our discussion below, as anticipated by instances of quantum bistable switching around the vacuum Rabi resonance in the trajectories of Figs. \ref{fig:Blockade}(b, c), which are absent from the resonance of Fig. \ref{fig:Blockade}(a).

\subsection{Emergence and persistence of photon blockade}
\label{subsec:emergpb}

In order to account for the thermodynamic limit with respect to photon blockade for a large $n$, we read in \cite{Carmichael2015} that an $n$-photon resonance occurs for $\Delta\omega=\pm g/\sqrt{n}$, while the ($n+1$)-photon resonance is ``blocked'' by an effective detuning $(\omega_{n+1,\pm}-\omega_{n,\pm})-\omega_{d} \approx \mp \kappa\sqrt{n_{\rm sc}/n}$. To arrive at this expression, we have used a Taylor expansion of the square root in $\pm g(\sqrt{n+1}-\sqrt{n}) \approx \pm g/(2\sqrt{n})$ \citep{spontdressedstate}, and substituted for $g$ from the definition of $n_{\rm sc}$. At the same time, Eq. \eqref{neocl} provides an indication on the extent of semiclassical nonlinearity through an offsetting detuning reading
\begin{equation}\label{detepskappa}
(\Delta\omega/g)^2=2(\varepsilon_d/\kappa)^2(\sqrt{1+(1/4)(\kappa/\varepsilon_d)^4}-1),
\end{equation}
to produce $|\alpha_{\rm ss}|^2=(\varepsilon_d/\kappa)^2$. Both regimes, that of photon blockade and that of neoclassical bistability, are extended and contrasted for $\kappa/\varepsilon_d\to 0, \kappa/g\to 0$, while for large values of the ratio $\varepsilon_d/\kappa$ we move towards the boundary $\Delta\omega/g=\pm\kappa/(2\varepsilon_d)$ where individual transitions are not resolvable, as we will discuss in Sec. \ref{sec:CABistanh}. Conversely, for $\varepsilon_d/\kappa \ll 1$ from Eq. \eqref{detepskappa} we obtain $\Delta\omega/g=\pm 1$, relevant to the build-up of the vacuum Rabi resonance we met in Sec. \ref{subsec:mappringKerr}. An appropriate choice for the scaled drive, reflecting the impact of nonlinearity in the strong-coupling limit, would then be $\sqrt{(\varepsilon_d/\kappa)^2/n_{\rm sc}}=2 \varepsilon_d/g$.

As $n_{\rm sc}$ is varied from $\sim 1$ to $\sim 10^6$ in frames (a-d) of Fig. \ref{fig:Blockade}, we find that the effective detuning does not compromise the resonances formed for successive values of $n$, while the regime of multi-photon blockade expands to include about nine visible resonance peaks whose location is determined predominantly by the discrete level structure. The Wigner functions drawn in the insets of frames (a, b) correspond to the peak of the vacuum Rabi resonance, displaying a mild nonlinearity with respect to the vacuum state. On the other hand, the Wigner functions in frames (c, d) evidence the onset of pronounced quantum fluctuations intertwined with the displayed low-photon nonlinearity \cite{QO2} in the region of the five-photon resonance with a peak expected at $\Delta\omega/g=1/\sqrt{5}\approx 0.4472$.

It then becomes clear that fluctuations do not vanish as $n_{\rm sc}$ is taken to infinity and $|\alpha_{\rm ss}|^2/n_{\rm sc}$ itself vanishes. Rather, they reveal conspicuously the discrete energy spectrum of the JC interaction. For growing $n$, the successive $n$-photon resonances become increasingly sharper, and their width approaches the order of magnitude of $\kappa$, where $\kappa/g\to 0$ (for $n_{\rm sc}\to \infty$). In this regime, low-amplitude bistability with large fluctuations accompanies the buildup of a multi-photon resonance, as we can see in the Wigner functions of Fig. \ref{fig:5P6PWigner}I(a-c) drawn for varying detuning when scanning across the sharp {\it post} saturation five-photon resonance peak {\it 5P}. The quantum trajectory for the intracavity field in frame (d) of Panel I shows quantum-fluctuation switching between a vacuum state \footnote{also frequently called {\it dim state}, as its photon average can be very small in comparison to that of the excited state yet nonzero.} and an excited state with mean amplitude of modulus squared $|\alpha|^2 \approx 1.5$, abiding with the prediction of Eq. \eqref{neocl}. The emergence of neoclassical bistability with growing $\varepsilon_d/g$ is also evident for a detuning where one would expect the peak of the six-photon resonance, $\Delta\omega/g=1/\sqrt{6} \approx 0.4082$, as depicted in Fig. \ref{fig:5P6PWigner}II. In the Wigner function of frame (d), for example, instead of the six-photon state, one finds an attractor with a mean amplitude of modulus squared $|\alpha|^2 \approx 2.7$, a value close to the upper branch of the mean-field bistability curve. 

\begin{figure}
\begin{center}
\includegraphics[width=0.5\textwidth]{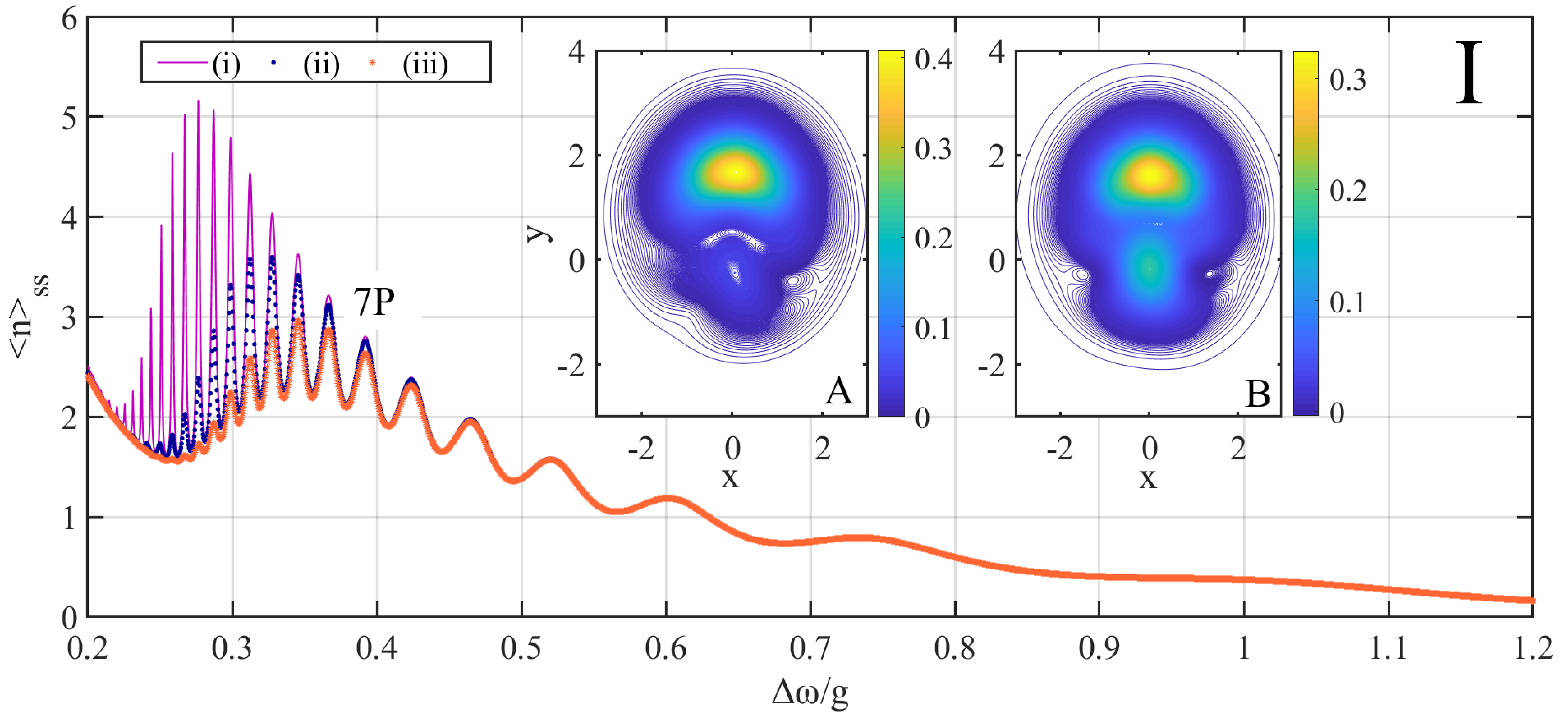}
\includegraphics[width=0.5\textwidth]{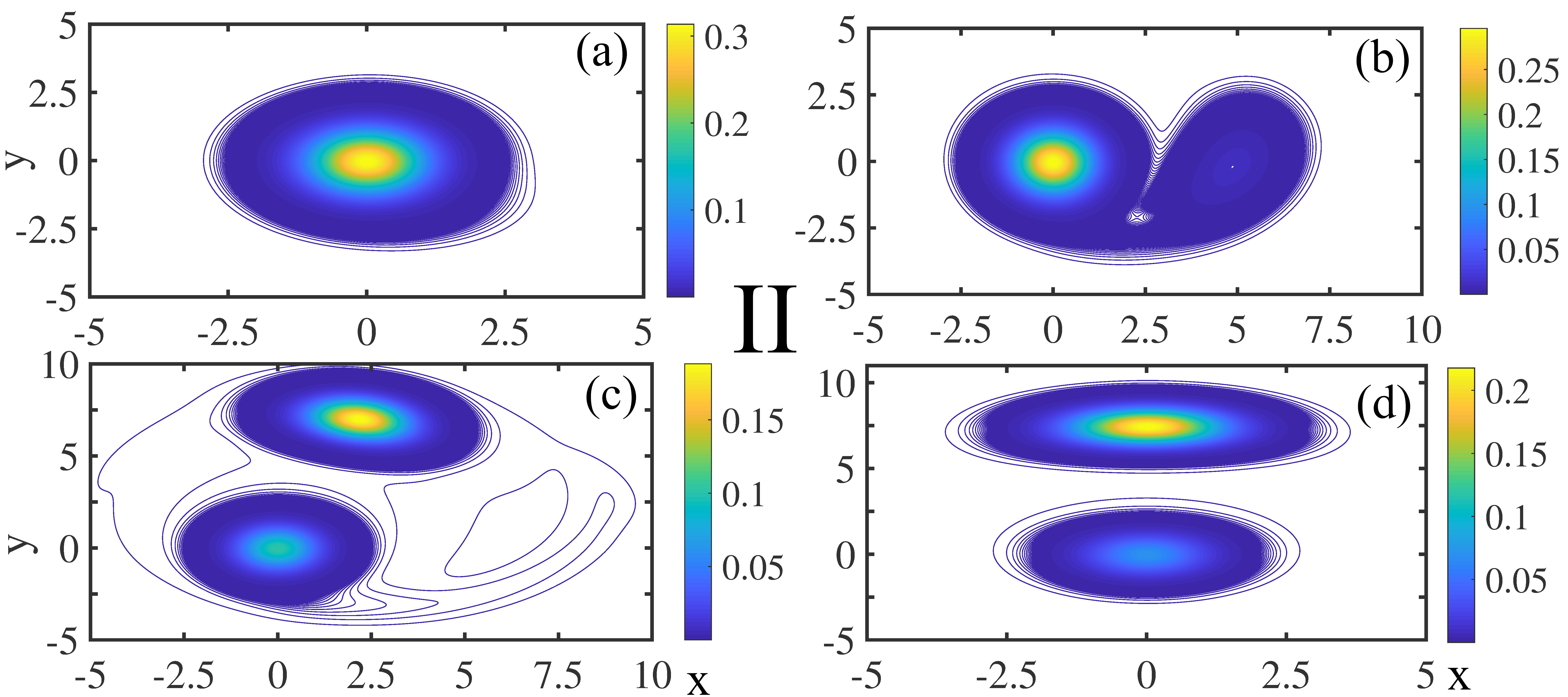}
\includegraphics[width=0.5\textwidth]{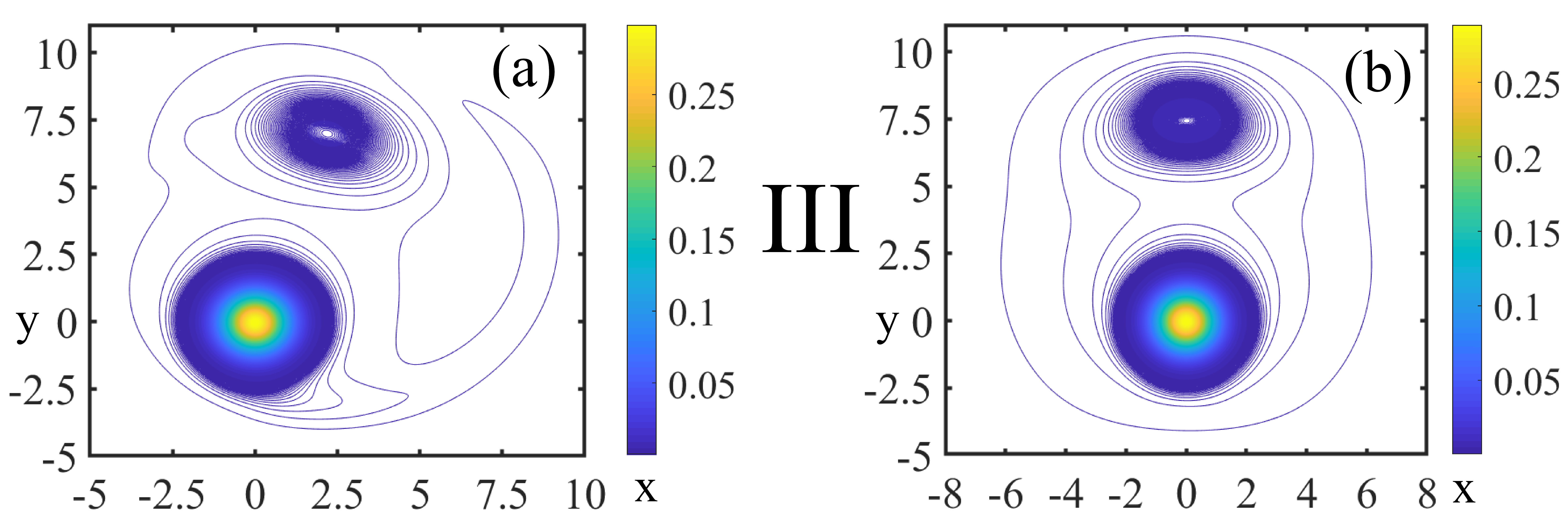}
\end{center}
\caption{{\it Complex-amplitude bimodality with higher photon numbers.} \underline{Panel I}: Steady-state average intracavity photon number obtained by the solution of the ME \eqref{MasterEq} for a fixed ratio $\varepsilon_{d}/g=0.18$, $\varepsilon_{d}/\kappa=900, 90, 45$, (and $g/\kappa=5000, 500, 250$) for the curves (i) [in solid purple], (ii) [in blue dots] and (iii) [in orange asterisks], respectively. The insets A, B depict the steady-state Wigner function $W(x+iy)$ for $\Delta\omega/g=0.392, 0.408$, respectively, along curve (iii). The corresponding seven-photon resonance is indicated by {\it 7P}. \underline{Panel II}: {\it Quasi}-probability distributions $Q(x+iy)$ of the intracavity field, obtained by the steady-state solution of the ME, for a fixed drive to coupling strength ratio $\varepsilon_d/g=0.245$, a fixed detuning to coupling strength ratio $\Delta\omega/g=0.1$, and a varying ratio $g/\kappa=10, 20, 10^2, 10^4$ in frames {\bf (a)}-{\bf (d)}, respectively. \underline{Panel III}: Steady-state {\it quasi}-probability distributions $Q(x+iy)$ of the intracavity field for $\varepsilon_d/g=0.245$, $\Delta\omega/g=0.1$ and $g/\kappa=10^2, 10^4$ in frames {\bf (a)}, {\bf (b)}, respectively, in the presence of spontaneous emission with rate $\gamma/\kappa=1$.}
\label{fig:channel}
\end{figure}

Bistability features in photon blockade in spite of the obvious disagreement between quantum dynamics and the mean-field response. In Panel I of Fig. \ref{fig:channel}, we keep $g$ and $\varepsilon_{d}$ constant, with $\varepsilon_{d}/g=0.18$, and increase $\kappa$ [in the spirit of Fig. \ref{fig:Blockade}] when plotting the three curves (i)-(iii) depicting the steady-state average photon number. Comparison between curve (i) of Fig. \ref{fig:channel}I and frame (d) of Fig. \ref{fig:Blockade} demonstrates that resonances become sharper and occur closer to the value $\Delta\omega/g=\pm 1/\sqrt{n}$ with decreasing $\varepsilon_{d}/g$, provided that $n_{\rm sc}$ is large enough. Curves (i)-(iii) share the same value of $\varepsilon_{d}/g$, double of that used for Fig. \ref{fig:Blockade}, with decreasing values of $n_{\rm sc}$. Their difference is appreciable only in the multi-photon resonances with $n\geq 7$, where a larger $n_{\rm sc}$ signifies sharper resonances. This is in contrast to Fig. \ref{fig:Blockade}, where such a disparity occurs for lower values of $n$ for the same values of $n_{\rm sc}$.

We then proceed to plot the Wigner functions of the intracavity field in the two insets of Panel I in Fig. \ref{fig:channel}, for a dip and the peak shaping the seven-photon resonance $7P$ along curve (iii). Quantum fluctuations organizing neoclassical bistability cannot secure a good representation of the vacuum state, as was the case in Panel I of Fig. \ref{fig:5P6PWigner}, resulting in softened features \footnote{Semiclassical bistability in Panel I of Fig. \ref{fig:channel} sets in for $\Delta\omega/g \approx 0.59$, past the first few resonances \textemdash{compare} with Fig. \ref{fig:Blockade}(d). We also note that for a given (nonzero) value of $\Delta\omega/g$ and a large enough $n_{\rm sc}$, the neoclassical bistability curve against $\varepsilon_d/g$ is not affected by a further decrease of the dissipation rate, as Eq. \eqref{modulusgeps} suggests for higher excitation.}. Returning to Fig. \ref{fig:Blockade}(d), we find that the enhancement of bimodality with respect to the varying representation of the vacuum and excited states across an $n$-photon resonance close to saturation can be seen as a competition between the JC interaction and coupling to the external field under vanishing dissipation. Therefore, in the thermodynamic limit and provided that $\varepsilon_d/\kappa$ is appreciably larger than unity, promoting $g$ exposes the quantum energy scale against the breakdown of photon blockade, while a growing $\varepsilon_{d}$ saturates fast the $n$-photon transition before it merges with the $(n+1)$-peak and eventually disperses into a continuum of unresolved resonances. This gradually brings us to the regime of large intracavity photon numbers approaching the order of a diverging $n_{\rm sc}$ as optical bimodality \textemdash{the} manifestation of a semiclassical nonlinearity that cannot be regarded as a negligible perturbation \textemdash{sets} in; this region is discussed in Sec. \ref{sec:CABistanh}. In that sense, the multi-photon transitions responsible for the nonlinear response acquire now a more continuous and classical dimension, in contrast to the regime of photon blockade where raising $n_{\rm sc}$ reveals more resonances as unabated quantum fluctuations.

\section{Spontaneous emission in the strong-coupling limit}
\label{sec:spontBl}

Before joining the two regions, we will make a brief detour to comment on the r\^{o}le played by spontaneous emission. Its inclusion adds the term $(\gamma/2)(2\sigma_{-}\rho \sigma_{+}-\sigma_{+}\sigma_{-}\rho-\rho\sigma_{+}\sigma_{-})$ on the right-hand side of ME \eqref{MasterEq} and breaks the conservation law for the pseudospin, introducing a new semiclassical scaling relation akin to a weak-coupling limit. The complex intracavity mean-field amplitude is written as \cite{Carmichael2015}
\begin{equation}\label{GammaOB}
\alpha_{\rm ss}=(\varepsilon_d/\kappa)\, h(\Delta\omega/\kappa, \tilde{\gamma}/\kappa, g^2/|\tilde{\gamma}|^2, |\alpha_{\rm ss}|^2/\tilde{n}_{\rm wc}),
\end{equation}
where 
\begin{equation*}
\begin{aligned}
&h(\Delta\omega/\kappa, \tilde{\gamma}/\kappa, g^2/|\tilde{\gamma}|^2, |\alpha_{\rm ss}|^2/\tilde{n}_{\rm wc})=\{1-i\Delta\omega/\kappa \\
&+ [g^2\tilde{\gamma}/(\kappa |\tilde{\gamma}|^2)](1+|\alpha_{\rm ss}|^2/\tilde{n}_{\rm wc})^{-1}\}^{-1},
\end{aligned}
\end{equation*}
with $\tilde{\gamma}\equiv \gamma/2+i\Delta\omega$ and $\tilde{n}_{\rm wc}\equiv|\tilde{\gamma}|^2/(2g^2)$. This produces the Maxwell-Bloch scaling law corresponding to absorptive optical bistability on resonance ($\Delta\omega/g=0$), as mentioned in the Introduction, together with the weak-coupling scaling parameter $n_{\rm wc}=\gamma^2/(8g^2)$ that we will make use of to quantify decoherence. For $\gamma=0$, it yields the scaling parameter $n_{\rm K}$ we have already met in Sec. \ref{subsec:mappringKerr} when making the semiclassical perturbative mapping to the Kerr oscillator, valid only for very low excitation. A detailed comparison between the quantum and semiclassical pictures in the strong-coupling limit with spontaneous emission present lies beyond the scope of our work. Nonetheless, we may note straight away that coupling of the two-level atom to the radiation modes of the reservoir has an ostensible impact on quantum coherence and the low-amplitude bistability associated with photon blockade. 
\begin{figure}
\begin{center}
\includegraphics[width=0.5\textwidth]{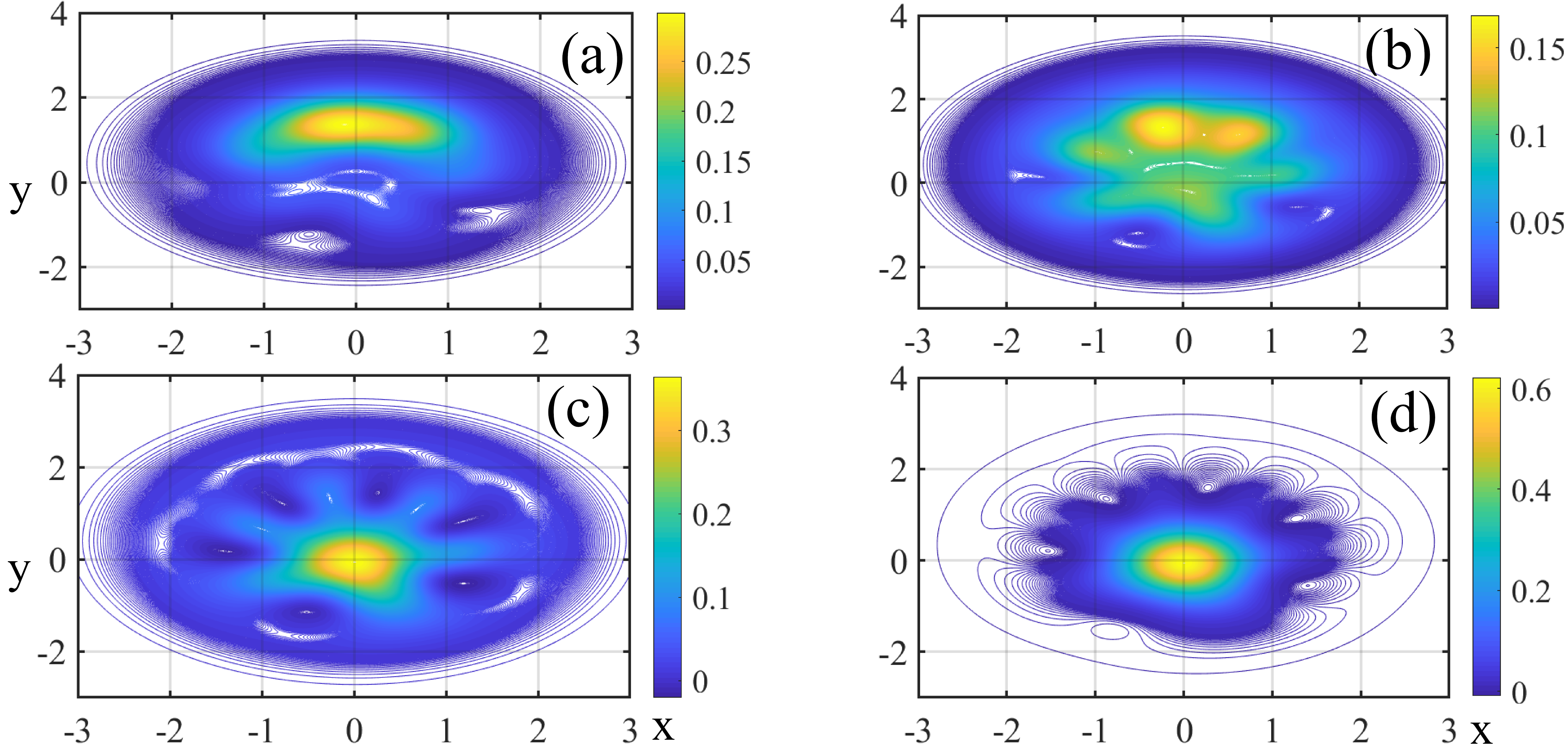}
\includegraphics[width=0.475\textwidth]{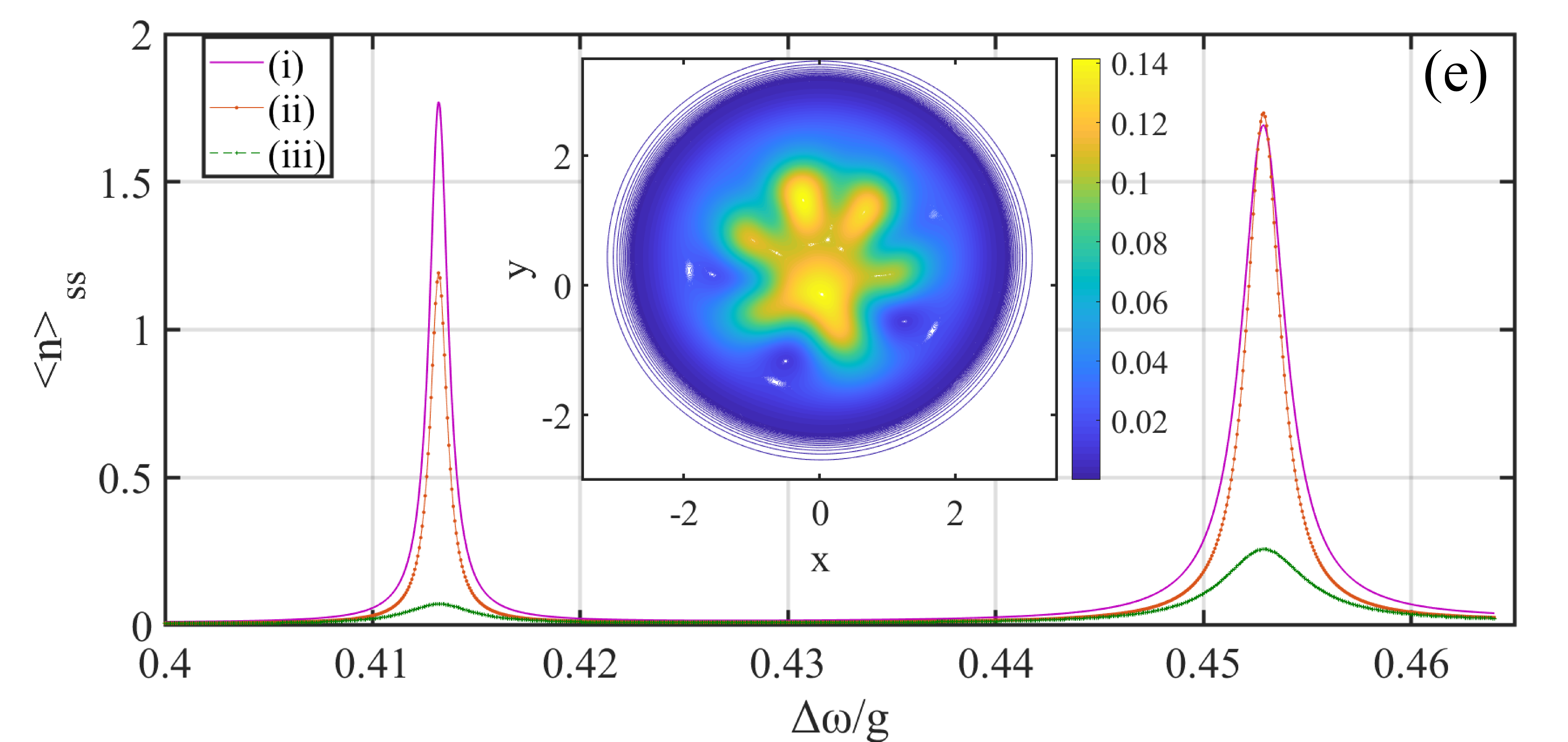}
\end{center}
\caption{{\it Spontaneous emission and photon blockade.} Wigner distributions $W(x+iy)$ of the intracavity field, obtained by the steady-state solution of the ME \eqref{MasterEq}, for a fixed detuning and drive to coupling strength ratio $\Delta\omega/g=0.4132$ and $\varepsilon_{d}/g=0.1$, respectively, in the presence of spontaneous emission with $\gamma/\kappa=0.2, 2, 20, 200$ in frames {\bf (a)}-{\bf (d)}, respectively. {\bf (e)} Steady-state intracavity photon number for variable detuning, obtained from the solution of Eq. \eqref{MasterEq} in the presence of spontaneous emission, for $\varepsilon_{d}/g=0.1$ and $\gamma/\kappa= 2, 20, 200$ in the curves (i) [solid purple], (ii) [orange dots], (iii) [green crosses], respectively. The five (six)-photon resonance peak is denoted by {\it 5P} ({\it 6P}). The inset depicts the Wigner distribution of the intracavity field for the same driving point as in frames (a-d) but with $\gamma/\kappa=3$. In all frames, $g/\kappa=5000$.}
\label{fig:GammaWigner}
\end{figure}

In frames (a-d) of Fig. \ref{fig:GammaWigner}, we follow the decay of the six-photon resonance peak {\it 6P} as $n_{\rm wc}$ is varied from $\sim 10^{-10}$ to $\sim 10^{-4}$. With increasing $\gamma/\kappa$, the $n$-photon resonances are compromised while their relative height changes to favor lower values of $n$ [frame (e)] as a result of the inability to sustain coherence over a many-photon process. More than the expected broadening of the quasi-probability distribution with pronounced quantum fluctuations, we observe a transition from the bright metastable state to the vacuum state as the photon occupation is lowered. This evidences once more the underlying connection of bistability and photon blockade; the peak height, which subsequently falls, is again in correspondence to the excited neoclassical state we met in Fig. \ref{fig:5P6PWigner} II(d). It is, however, quantum fluctuations that efface bistability in a manner distinct to merely altering the switching rate between two metastable states, as dictated by the semiclassical nonlinear dynamics. We will take this discussion further in the next section. 

\section{From complex-amplitude bimodality to the anharmonic-ladder oscillator}
\label{sec:CABistanh}

We will now turn to the higher cavity-field amplitudes, following the breakdown of photon blockade by means of a first-order quantum phase transition \cite{Carmichael2015}. Here quantum fluctuations induce switching between states differing by a substantial amount of photon occupation. We can get a good impression of the approach to the strong-coupling limit from the {\it quasi}-probability distributions $Q(x+iy)$ of the intracavity field, as depicted in Panel II of Fig. \ref{fig:channel}. In this region of the drive parameter space, quantum fluctuations essentially restore the neoclassical nonlinearity, with a development of complex-amplitude bimodality for increasing $n_{\rm sc}$. In frames (a-d), we witness the birth of a highly-excited metastable state out of the vacuum state in the $Q$ function for the intracavity field, which eventually assumes most of the excitation probability following a spiral rotation of the mean complex amplitude in the first quadrant of the phase space. 

When $4 g^2|\alpha_{\rm ss}|^2 \gg \Delta\omega^2$, i.e., for the reverse asymptotic behavior to the one considered in Sec. \ref{subsec:mappringKerr}, we find two roots of the neoclassical equation, 
\begin{equation}\label{modulusgeps}
|\alpha_{\rm ss, \pm}|^2\approx [(g\pm 2\varepsilon_d)/(2\Delta\omega)]^2,
\end{equation}
as $\kappa \to 0$, while there is no mean-field bistability for $\varepsilon_d > g/2$. Although $|\alpha_{\rm ss, +}|$ in Eq. \eqref{modulusgeps} sets an upper boundary for the cavity excitation (see e.g. the successive excited metastable states in Fig. 3(a) of \cite{StabilityFSC} for increasing $g/\kappa$), it is still up to the quantum fluctuations to determine the balance between the two metastable states \textemdash{a} situation pictured in frame (d) of Panel II \textemdash{and} consequently the value of $\braket{a^{\dagger}a}_{\rm ss}$. What is perhaps more illuminating regarding the coherent superposition of the drive field with amplitude $\varepsilon_d$ and the field re-radiated by the two-level atom, with amplitude $g\braket{\sigma_{-}}_{\rm ss, +} \to g/2$, is the other root of the neoclassical equation in the limit under consideration, $|\alpha_{\rm ss, -}|^2\approx [(g-2\varepsilon_d)/(2\Delta\omega)]^2$, corresponding to the unstable branch of the bistability curve \textemdash{a} destructive interference between the two fields. This expression produces the form $0/0$ on resonance at the critical point. At the ratio $g/\kappa \approx 43$, a value between $20$ and $100$ [in frames (b) and (c) respectively], we cross the boundary of the first-order dissipative quantum phase transition; for the corresponding value of $n_{\rm sc}=n_{\rm sc, b}$ as determined by quantum fluctuations, the peak heights of the two states are equal in the $Q$ function. On approaching resonance, $\Delta\omega/g \to 0$, the two states form an angle $\theta_{\pm} \approx \pm \sqrt{n_{\rm sc}/|\alpha_{\rm ss}|^2}$ with respect to the $x$-axis for $2 \varepsilon_d/g \gg 1$, a limit where the JC interaction can be treated as a perturbation [see Eq. \eqref{thetass} in Sec. \ref{sec:symmbr}].
\begin{figure}
\begin{center}
\includegraphics[width=0.5\textwidth]{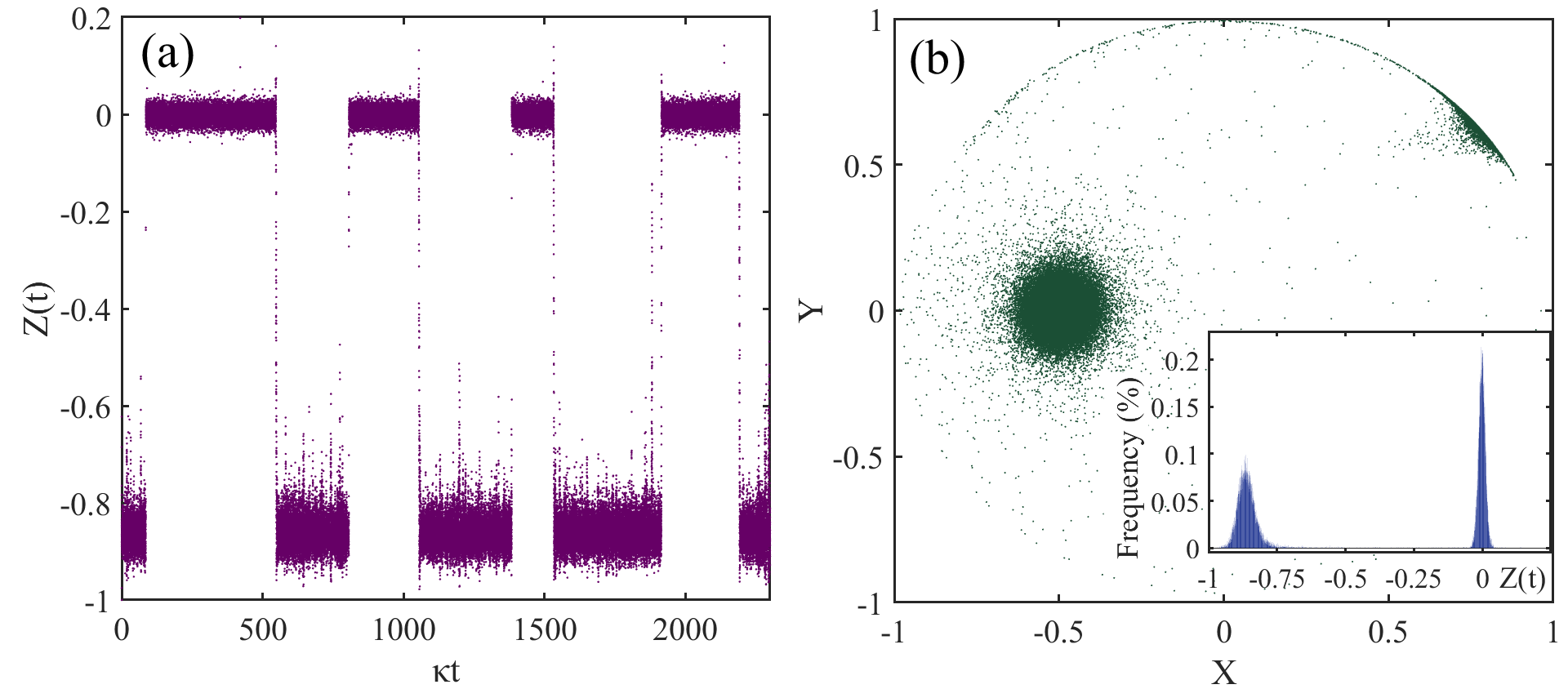}
\end{center}
\caption{{\it On the boundary of a first-order dissipative quantum phase transition.} Sample quantum trajectory depicting $Z(t) \equiv \braket{\sigma_z(t)}$ in frame {\bf (a)} and the projection of atomic coherence (as a distribution) on the equatorial plane of the Bloch sphere [with $(X,Y)(t) \equiv \braket{\sigma_{(x,y)}(t)}$] in frame {\bf (b)}. The inset depicts a frequency histogram (with a large bin number) for the inversion $Z(t)$, as extracted from the sample trajectory. Here, $g/\kappa=43.48$, with the remaining parameters being identical to those used for Panel II of Fig. \ref{fig:channel}.}
\label{fig:qubitqpt}
\end{figure}
In Fig. \ref{fig:qubitqpt}, we depict quantum-fluctuation switching between equiprobable metastable states \footnote{The qualification of equal probability is used in a loose sense here, since the condition of equal areas under the two peaks in the {\it quasi}-probability distribution does not also mean equal peak heights in the presence of quantum fluctuations for a finite-size scaling \textemdash{see} Sec. VIII of [Bonifacio R. {\it et al.}, Phys. Rev. A {\bf 18}, 2266 (1978)] considering as well the weak-coupling limit of many-atom absorptive resonant bistability (with fluctuations scaled by the number of atoms in the cavity).} in a sample realization for the atomic inversion next to the coherence $2\braket{\sigma_{-}(t)}$ for $n_{\rm sc, b}$. The projection of the distribution corresponding to the excited state is bounded by the circumference of the circle $|2\braket{\sigma_{-}(t)}|=1$ on the equatorial plane, in the limit $4 g^2|\alpha_{\rm ss}|^2 \gg \Delta\omega^2$, while the projection corresponding to the vacuum state lies inside the circle, as pictured in frame (b) [see also Eq. (17) of \cite{Carmichael2015}]. On resonance, the two equiprobable distributions are instead symmetrically positioned on the equatorial plane.  

The multi-photon resonances we met in Sec. \ref{subsec:emergpb} are smeared out as the intracavity amplitude grows with vanishing $\kappa$, and eventually merge into the ``tail-envelope'' of a split Lorentzian peaking at $\Delta \omega/g=\pm \kappa/(2\varepsilon_{d})$ in the direction of resonance, with $|\alpha_{\rm ss}|^2=(\varepsilon_{d}/\kappa)^2$. The effective detuning contains $\Delta\omega$ as a free variable of the split Lorentzian function, with $(\omega_{n+1,\pm}-\omega_{n,\pm})-\omega_{d} \approx -\Delta\omega\pm g/(2\sqrt{n})$, and is not considered in reference to the discrete steps along the two ladders for which the scaling relationship of photon blockade was formulated. In that nonlinear response, the disparity between neoclassical theory and quantum dynamics recedes with increasing $\varepsilon_d/\kappa$ (see, e.g., Fig. 4 of \citep{Carmichael2015}). 

Prompted by the emerging agreement between the two pictures and leaving bimodality aside for the moment, we set $\Delta\omega=g/\sqrt{\braket{n}}$ in Eq. \eqref{neocl} [hence selecting the upper excitation ladder without loss of generality, while retaining the semisclassical nonlinearity in the response (which is canceled at $\Delta\omega=g/\sqrt{2\braket{n}}$)], with $\braket{n} \equiv \braket{a^{\dagger}a}_{\rm ss}$ treated as a continuous variable. This yields
\begin{equation}\label{UpperOsc}
\braket{n} + (g/\kappa)^2[1-\braket{n}(4\braket{n}^2+1)^{-1/2}]^2=(\varepsilon_{d}/\kappa)^2,
\end{equation}
giving $\braket{n} \approx (\varepsilon_{d}/\kappa)^2-n_{\rm sc}=n_{\rm sc}[(2\varepsilon_{d}/g)^2-1]$ above threshold on approaching the thermodynamic limit with high excitation \textemdash{a} benchmark of the empty-cavity excitation with respect to the scaling photon number. At the mean-field level, this result brings us to the model of the quantum $\sqrt{n}$ anharmonic oscillator on resonance, formulated via either of the two excitation ladders of the JC spectrum (see also Sec. 16.3.5 of \citep{QO2}). As we will see in Sec. \ref{sec:symmbr}, however, the r\^{o}le of quantum fluctuations is special about the critical point of the JC second-order quantum phase transition. The presence of semiclassical bistability for a low-photon cavity occupation has already been experimentally ascertained in \cite{remnants} (see e.g. Figs. 2 and 3 therein), while the discrete nature of the JC spectrum underlies the stability of the two metastable states (vacuum and excited) on approaching the thermodynamic limit, as demonstrated in \cite{StabilityFSC} for large photon numbers.

Once more, spontaneous emission has a drastic impact on the system response. The average photon number in the steady state drops from the value $\braket{n}_{\rm ss}\approx 36.4, 42.8$ to $\braket{n}_{\rm ss}\approx 2.2, 3.7$, for $n_{\rm wc} \sim 10^{-5}, 10^{-9}$, respectively, as we see when comparing frames (c,d) of Panel II to frames (a,b) of Panel III in Fig. \ref{fig:channel}. The excited state is only rarely visited in the course of quantum-fluctuation switching, while the boundary of a first-order dissipative quantum phase transition cannot be crossed for the detuning and drive strength under consideration.  The excited state remains in place in the phase space for varying $\gamma/\kappa$ at a given point in the drive parameter space; what changes is its participation in the cavity output. This is in contrast to what happens in photon blockade, a regime where the entire landscape of quantum dynamics is very prone to decoherence, as we have seen in Sec. \ref{sec:spontBl}.

\section{Symmetry breaking and critical behavior}
\label{sec:symmbr}

\begin{figure}
\begin{center}
\includegraphics[width=0.5\textwidth]{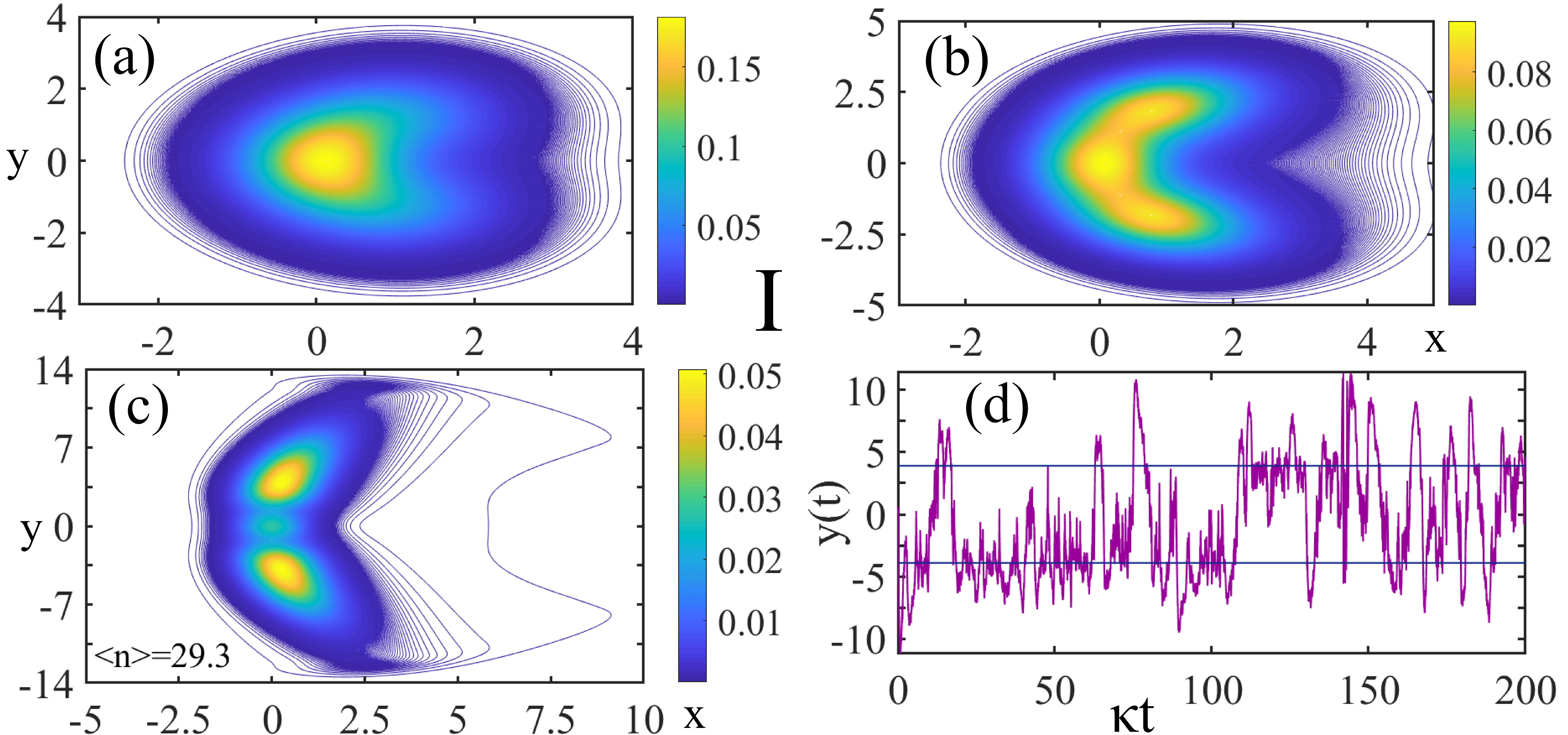}
\includegraphics[width=0.5\textwidth]{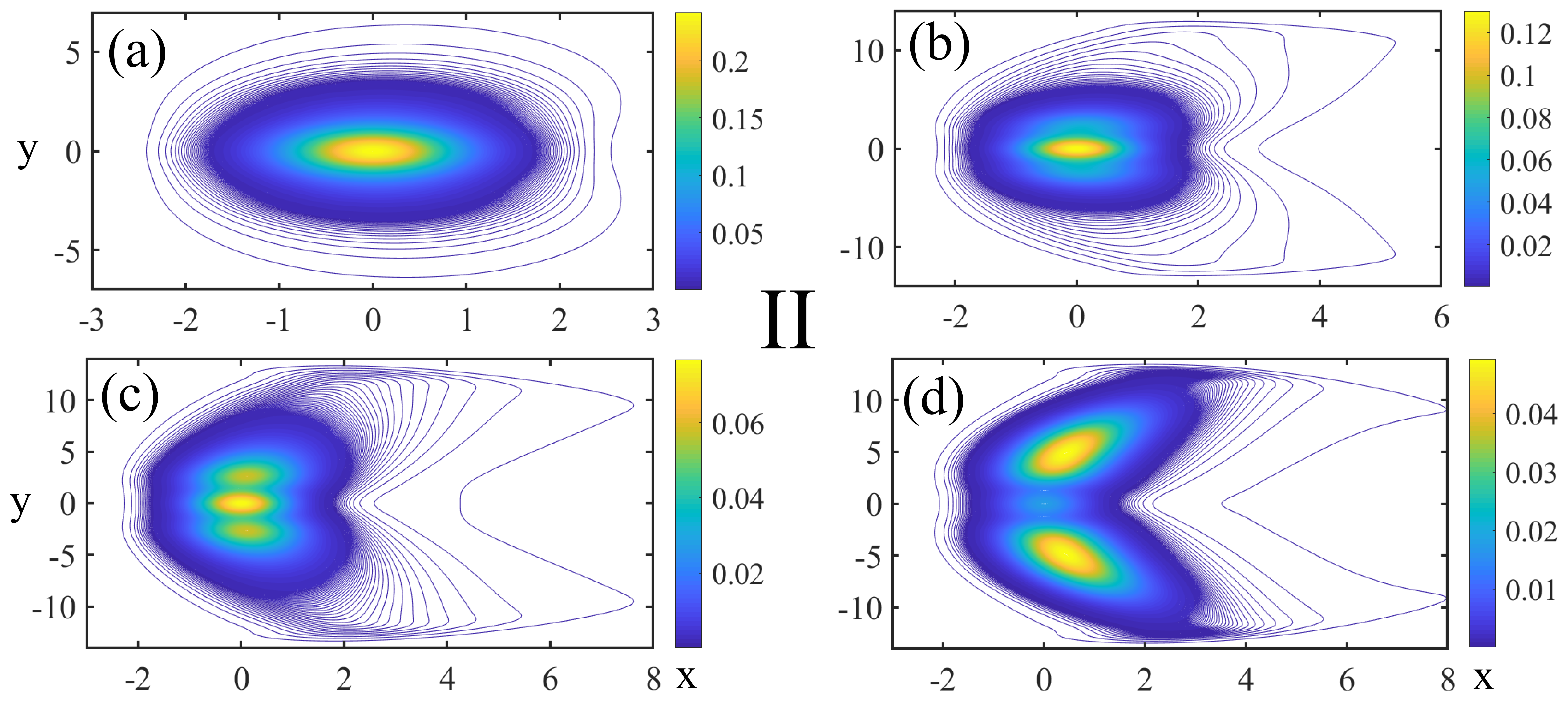}
\end{center}
\caption{{\it Quantum phase bimodality below threshold.} \underline{Panel I}: {\it Quasi}-probability distributions $Q(x+iy)$ of the intracavity field, obtained by the steady-state solution of the ME, for a fixed drive to coupling strength ratio $\varepsilon_d/g=0.495$, a fixed detuning to coupling strength ratio $\Delta\omega/g=0$ (resonance), and a varying ratio $g/\kappa=5, 10, 100$ in frames {\bf (a)}-{\bf (c)}, respectively. In frame {\bf (d)} we plot a sample trajectory depicting $y(t)={\rm Im}[\braket{a(t)}]$ for the same parameters as in frame (c). The two horizontal lines mark the imaginary part of the two positions in the phase space where the $Q$ function of frame (c) peaks. \underline{Panel II}: Steady-state $Q$ functions at resonance for $g/\kappa=100$ and an increasing ratio $\varepsilon_d/g=0.4, 0.46, 0.48, 0.498$ in frames {\bf (a)}-{\bf (d)}, respectively.}
\label{fig:BelowthrPB}
\end{figure}

The two coherent interactions compete with each other much more clearly on resonance (see also the opening paragraphs in Sec. 16.3.2 of \citep{QO2}), where the neoclassical equations of motion yield a steady-state solution with the typical bifurcation of a second-order phase transition \citep{spontdressedstate, QO2}. Quantum fluctuations organize attractors about the complex-conjugate neoclassical states and realize phase bistability. Above threshold, with $\varepsilon_{d}/g \geq 1/2$, the complex amplitude of the intracavity field is given by the pair of states \cite{spontdressedstate}
\begin{equation}\label{abovethrneocl}
\alpha_{\rm ss}=(\varepsilon_d/\kappa)\{1 - [g/(2\varepsilon_d)]^2\} \pm i\,[g/(2\kappa)]\sqrt{1-[g/(2\varepsilon_d)]^2},
\end{equation}
which shows that cavity occupation grows without bounds as $g/\kappa \to \infty$, following the law (from Eq. \eqref{neocl} on resonance \cite{Carmichael2015}) 
\begin{equation}\label{abovethr}
|\alpha_{\rm ss}|^2=[g^2/(4\kappa^2)]\,[(2\varepsilon_{d}/g)^2-1],
\end{equation}
something we have already met as an asymptotic form for an intensity-dependent detuning approaching zero with increasing intracavity excitation. The neoclassical equations predict a zero intracavity amplitude at the critical point. In fact, quantum fluctuations produce once more a visible deviation from the mean-field prediction, as we can see in Fig. \ref{fig:BelowthrPB}; on approaching the strong-coupling limit, a bimodal distribution with a long stretch in the phase space emerges from the vacuum state below threshold \textemdash{corroborated} by the switching trajectory drawn in frame (d) of Panel I for $n_{\rm sc}=2.5\times 10^3$, with $\braket{n}_{\rm ss} \approx 29$ [see also the Wigner distribution of Fig. 16.12(c) in \cite{QO2}, at the critical point for $n_{\rm sc}=25$]. In Panel II, we track the development of quantum-fluctuation phase bimodality for growing $\varepsilon_d/g$ below threshold, a precursor to the decay of the unstable vacuum state at a mean-field bifurcation. The two pictures are coming closer together for increasing $\varepsilon_d/g$ above threshold \footnote{We note that in the limit of weak excitation within a dressed-state formulation, where $\varepsilon_d/g \ll 1$, the photon number average differs from zero by $\braket{a^{\dagger}a}_{\rm ss} \approx 2 (\varepsilon_d/g)^4$ \textemdash{see} Eq. (16.203) of \cite{QO2}.}.
\begin{figure*}
\begin{center}
\includegraphics[width=0.5\textwidth]{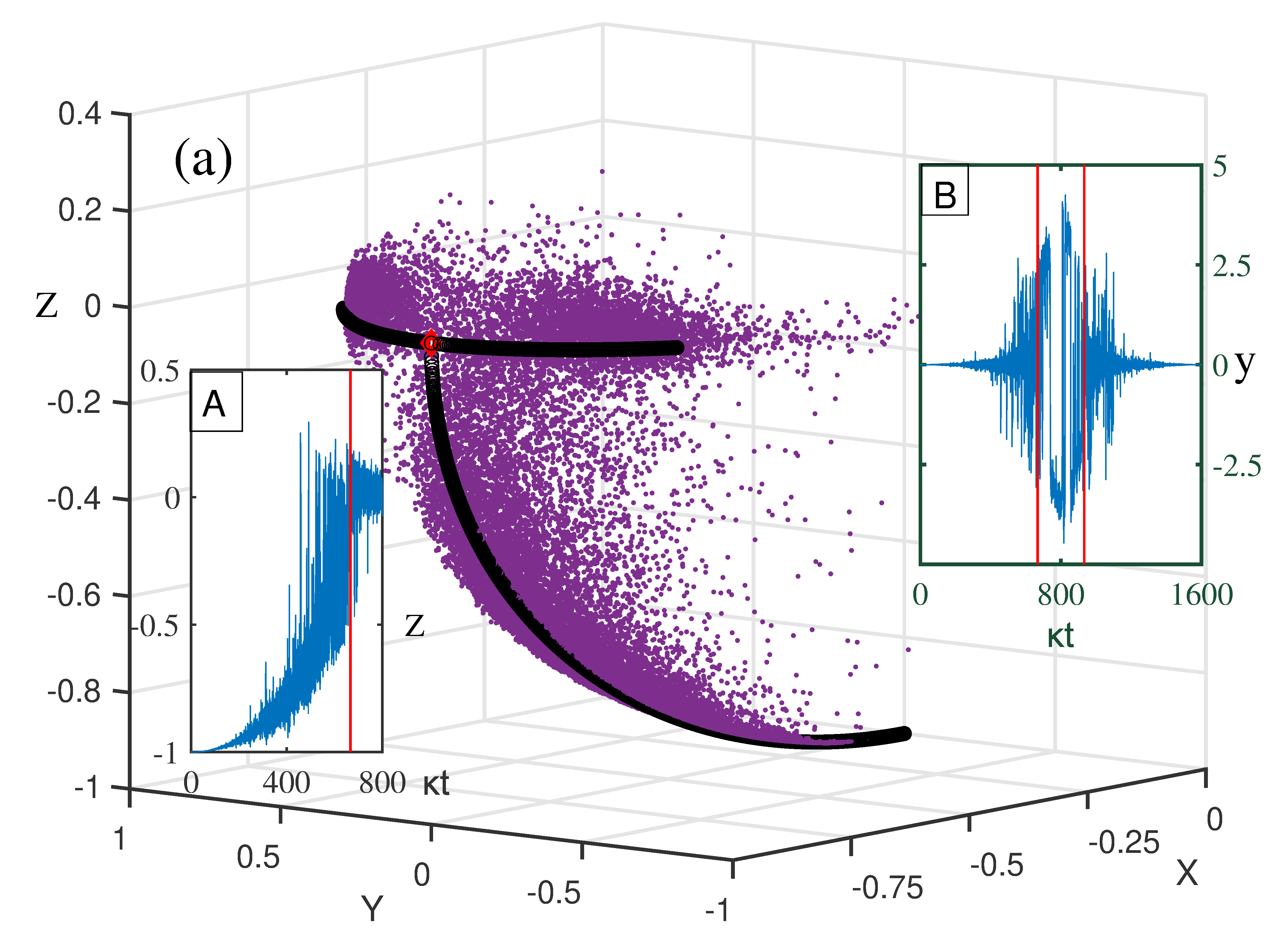}
\includegraphics[width=0.48\textwidth]{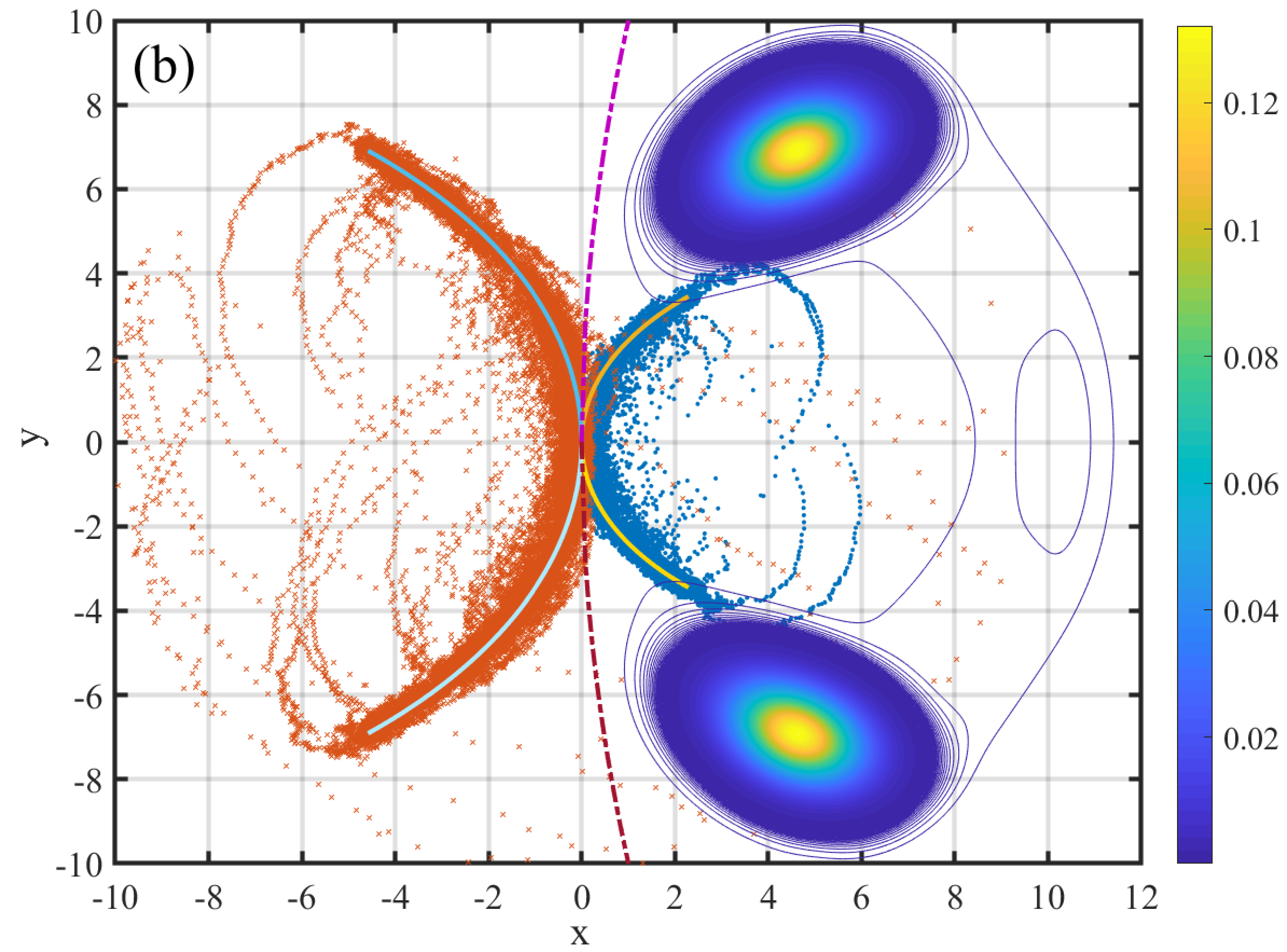}
\end{center}
\caption{{\it Symmetry breaking in the quantum picture.} Time-dependent averages from single quantum trajectories with variable driving to coupling strength ratio, $\varepsilon_{d}(t)/g$, ranging from zero to $0.6$ in a linear fashion, and backwards. The overall (in both directions) dimensionless scan time is $\kappa T=1.6 \times 10^3$. {\bf (a)} Bloch sphere representation of the two-level polarization averages $\braket{\sigma_{i}(t)}$ ($i=x, y, z$), with $(X(t), Y(t), Z(t))\equiv (\braket{\sigma_{x}(t)}, \braket{\sigma_{y}(t)}, \braket{\sigma_{z}(t)})$ and initial condition $(X(0), Y(0), Z(0))=(0,0,-1)$ for $g/\kappa=12.5$. The curves in black depict $\braket{\boldsymbol{\sigma}(\varepsilon_{d}/g)}$, calculated from the zero quasi-energy states (see Fig. 2 of \cite{DiracJG2018}), and the red diamond marks the critical point. The inset A shows the time evolution of $Z(t)\equiv \braket{\sigma_z(t)}$ while the inset B shows $y(t)\equiv{\rm Im}[\braket{a(t)}]$. The vertical lines in both insets demark the region above threshold [where $0.5< \varepsilon_{d}(t)/g \leq 0.6$]. {\bf (b)} Phase-space representation of the intracavity amplitude $\braket{a(t)}\equiv x(t)+iy(t)$ for two different values of $g/\kappa$: $12.5$, in blue dots; $25$, in orange crosses. The latter distribution, which corresponds to the superimposed contour plot of the $Q$ function for $\varepsilon_d/g=0.6$, is shown inverted with respect to the center of co-ordinates for clarity of presentation. Solid lines depict the corresponding neoclassical results above threshold from Eq. \eqref{abovethrneocl}. The bold dashed curves depict the two neoclassical branches for $g/\kappa=200$ with $0.5 \leq \varepsilon_{d}/g \leq 0.5 \sqrt{100/99}$, i.e., a scaled drive amplitude reaching only $0.5\%$ above threshold.}
\label{fig:phasebist}
\end{figure*}

In \cite{DiracJG2018}, we read that ``without dissipation, once the system is driven above the critical point, the field radiated by the two-level system is no longer strong enough to interfere destructively with the coherent drive.'' We build upon this remark in Fig. \ref{fig:phasebist}, where the drive amplitude is scanned from zero to a value above threshold and then back to zero in the course of individual realizations for two different values of $n_{\rm sc}$. We first visualize the organizing point and the onset of the second-order dissipative quantum phase transition in frame (a). During the stochastic evolution, the Bloch vector \textemdash{positioned} initially at the south pole of the sphere \textemdash{fluctuates} about the neoclassical curve defined by $X(t)\equiv\braket{\sigma_{x}(t)}=-2\varepsilon_{d}(t)/g$ and $Z(t)\equiv\braket{\sigma_{z}(t)}=-\sqrt{1-[2\varepsilon_{d}(t)/g]^2}$ and, after meeting the critical point, follows two separating symmetric distributions positioned about the equatorial plane above threshold, as inset A shows. 

The sample trajectory in frame (a) of Fig. \ref{fig:phasebist} follows closely the average polarization $\braket{\boldsymbol{\sigma}(\varepsilon_{d}/g)}\equiv\braket{\phi_0|\boldsymbol{\sigma}|\phi_0}$ in the zero quasi-energy eigenstate $\ket{\phi_0}$ (with dissipation {\it a priori} absent), which captures the symmetry breaking in alignment with the prediction of the neoclassical theory in the presence of (driving and) dissipation \cite{DiracJG2018}. We also note that fluctuations intensify significantly prior to attaining the critical point, in contrast to the response of the $\sqrt{n}$ anharmonic oscillator, as we can observe in the imaginary part of the coherent-state amplitude of the resonant cavity mode, drawn in inset B. The imaginary part of the intracavity amplitude fluctuates weakly in the beginning about the neoclassical state $\alpha_{\rm ss}=0$ \cite{spontdressedstate} before alternately switching between the two attractors determined eventually by Eq. \eqref{abovethrneocl} past the critical point and beyond the region defined by the critical behavior we met in Fig. \ref{fig:BelowthrPB}. These mean-field attractors are the states drawn by the solid lines in the phase-space distribution of frame (b). Their deviation with respect to the $x$-axis in the phase space is a monotonically decreasing function of $\varepsilon_d/g$, with an angle 
\begin{equation}\label{thetass}
\theta_{\pm}= \pm \arctan\{[(2\varepsilon_d/g)^2-1]^{-1/2}\}.
\end{equation}
This expression gives $\theta_{\pm} \approx \pm g/(2\varepsilon_d)$ in the limit $2\varepsilon_d/g \gg 1$. In the same limit, the Bloch vector fluctuates maximally between the attractors $(X=0, Y=\pm 1)$ on the equatorial plane. 

When quantum fluctuations are taken into account, the separation between the two distributions of phase bimodality in the phase space of the cavity field decreases when lowering $n_{\rm sc}$; in the extreme case where $g/\kappa \approx 1$, they merge to produce eventually a single squeezed coherent state in the steady-state response. In this case, squeezing of fluctuations occurs along the real axis, which is the direction of the neoclassical excitation amplitude $\varepsilon_d/\kappa$; this amplitude need not be large.  

Above threshold, quantum fluctuations are positioned around the neoclassical curves, more clearly so for growing $\varepsilon_d/g$. The discrete quasi-energy spectrum has now collapsed and is determined solely by the continuous variation of the drive amplitude \cite{QO2}. Hence, with increasing the value of $n_{\rm sc}$ we attain higher and higher cavity photon numbers for the same ratio of $\varepsilon_d/g$, scaling once more as $|\alpha_{\rm ss}|^2=(\varepsilon_{d}/\kappa)^2$ for $2\varepsilon_d/g \gg 1$, as predicted by Eq. \eqref{abovethr}. The coupling strength $g$ is manifestly absent in this limit, testifying to the vanishing interference with the field re-radiated by the two-level atom \textemdash{a} sharp contrast with the formation of multi-photon resonances in the region of photon blockade.

On approaching the thermodynamic limit, the fluctuations follow suit to stabilize the mean-field states of phase bimodality against the neoclassical prediction of nonstable limit cycles above threshold \cite{spontdressedstate}, as we can see in frame (b) of Fig. \ref{fig:phasebist}. Moreover, for $2\varepsilon_d/g \gg 1$, deviations due to quantum fluctuations in the $Q$ representation from the attractors comprising the bimodal distribution in the steady state are constrained inside a ring delimited by a coherent-state uncertainty, as opposed to the replication of a bifucation in the phase space that we met in Fig. \ref{fig:BelowthrPB} when accessing the region below the critical point. Those fluctuations also appear symmetric in the phase space and are squeezed in the radial direction, with the radius itself defined by the large coherent-state amplitude $\varepsilon_d/\kappa$. Extending finally to the generalized Jaynes-Cummings-Rabi model, we note that the many-photon strong-coupling limit obviates the need of employing many two-level emitters to attain high excitation; strong coupling to one emitter is sufficient \cite{QPTJCRabi}.

\section{Concluding discussion}

In conclusion, we have explored the definition of a strong-coupling thermodynamic limit for the driven dissipative JC interaction. This limit probes the ability of the field radiated by the two-level system, the fundamental element introducing the nonlinearity, to interact coherently with a resonant mode driven by a fixed, externally-imposed drive, when at the same time this coherent interaction is compromised by dissipation. We have shown that a strong coupling between the atom and its radiated field, resonating in the cavity, underlies the manifestation of photon blockade in the parameter regime where $\varepsilon_d \ll g \sim |\Delta\omega|$. Distinct resonance peaks mark the formation of effective two-level complexes with varying participation in the response as we attain the strong-coupling limit. The signal leaving the cavity comprises the individual photons taking part in this coherent process rather than the classical drive field, detuned off-resonance, despite the fact that the scaled semiclassical amplitude vanishes. As we access the region $\varepsilon_d \sim g \gg |\Delta\omega|$, we have demonstrated that the drive and re-radiated fields are progressively placed on equal footing, shaping the picture of bimodality in the phase space. Instead of the individual multi-photon resonances there is now a split-Lorentzian response \textemdash{evidence} of the ``collective'' form the nonlinearity now assumes \textemdash{with} peaks at detunings approaching a zero value for vanishing dissipation, which is where symmetry breaking takes place.

Moving now to resonance, $\Delta\omega=0$, below threshold the two fields interfere destructively to produce a zero neoclassical intracavity amplitude. On the other end, when driving above threshold with $2\varepsilon_d/g \gg 1$, the parameter $n_{\rm sc}$, scaling the cavity excitation to very large values, accounts only for a correction to the mean-field cavity output with respect to the empty-cavity response. In the region of the critical point, however, we have demonstrated the presence of significant quantum fluctuations against the neoclassical response, heralding symmetry breaking and the associated bifurcation; the ME predicts a bimodal steady-state distribution below threshold with an appreciable photon-number average in the strong-coupling limit. For high excitation above threshold, we can invoke the two excitation ladders of the JC spectrum, although in this case separately driven and with a square root of the corresponding bosonic number operator replacing the atom-field coupling term in the JC Hamiltonian. Fluctuations induce switching between these two ladders, conforming to mean-field phase bistability.  

In fact, bistability provides the connective net between the aforementioned frames, some times rather loose and others more palpable; we encounter it from the weak semiclassical deflection of the vacuum Rabi resonance up to the phase-bistable switching at resonance involving potentially innumerable photons. We have demonstrated, however, that quantum fluctuations are not there to merely realize of forecast mean-field bistability: the approach of a strong-coupling thermodynamic limit in photon blockade entails uncompromised resonances in continuing disagreement between the quantum picture and the semiclassical treatment. As we have also observed, among the two decoherence channels, cavity decay is not as effective in inducing switching between the JC excitation ladders as spontaneous emission, which justifies the special place reserved for the neoclassical equations and the pertinent scaling law. 

With the above remarks and qualifications in mind, sending the strong-coupling scale parameter $n_{\rm sc}$ to infinity exposes the underlying JC energy spectrum in either its discrete or continuous manifestation, depending on the relative weight of the two coherent coupling strengths. Quantum fluctuations determine the particulars of critical behavior, as captured by the experiment, based on the interplay between the coherent and incoherent interactions.  

\begin{acknowledgments}
I am grateful to H. J. Carmichael for his unfaltering guidance, and to J. Larson for several inspiring discussions. This work was supported by the Swedish Research Council (VR) as well as by the Knut and Alice Wallenberg foundation (KAW).
\end{acknowledgments}

\bibliography{SCL_paperB}

\end{document}